\documentclass[conference]{IEEEtran}
\IEEEoverridecommandlockouts
\usepackage{cite}
\usepackage{amsmath,amssymb,amsfonts}
\usepackage{textcomp}
\usepackage[a4paper, total={184mm,239mm}]{geometry}
\usepackage[utf8]{inputenc} %
\usepackage[T1]{fontenc}    %
\usepackage[table,xcdraw]{xcolor}
\usepackage{pifont}
\usepackage[colorlinks,
    linkcolor=red,citecolor=citecolor,urlcolor=blue]{hyperref}       %

\usepackage{graphicx}
\usepackage{xcolor}
\usepackage{todonotes}

\usepackage{amssymb}
\usepackage{threeparttable}
\usepackage{multirow}
\usepackage{amsmath}
\usepackage{bm}
\usepackage{float}
\usepackage{amsthm}
\usepackage{soul}
\usepackage[noend]{algpseudocode}
\usepackage{enumitem} %
\usepackage[subrefformat=parens,labelformat=parens]{subfig}

\usepackage{xspace}

\usepackage{multirow}

\definecolor{citecolor}{RGB}{34,139,34}
\definecolor{mydarkblue}{rgb}{0,0.08,1}
\definecolor{mydarkgreen}{rgb}{0.02,0.6,0.02}
\definecolor{mydarkred}{rgb}{0.8,0.02,0.02}
\definecolor{mydarkorange}{rgb}{0.40,0.2,0.02}
\definecolor{mypurple}{RGB}{111,0,255}
\definecolor{myred}{rgb}{1.0,0.0,0.0}
\definecolor{mygold}{rgb}{0.75,0.6,0.12}
\definecolor{myblue}{rgb}{0,0.2,0.8}
\definecolor{mydarkgray}{rgb}{0.,0.2,0.2}

\definecolor{lightred}{RGB}{255,235,235}
\definecolor{lightgreen}{RGB}{235,255,235}
\definecolor{lightblue}{RGB}{235,235,255}
\definecolor{lightcyan}{RGB}{235,255,255}
\definecolor{lightmagenta}{RGB}{255,235,255}
\definecolor{lightyellow}{RGB}{255,255,235}

\definecolor{qxkcolor}{RGB}{215,235,255}
\definecolor{softmaxcolor}{RGB}{230,235,255}
\definecolor{probxvcolor}{RGB}{255,255,235}

\definecolor{topkcolor}{RGB}{255,235,235}
\definecolor{zecolor}{RGB}{255,255,235}
\definecolor{dynacolor}{RGB}{235,255,255}

\definecolor{reviewcolor}{RGB}{0,0,200}

\newcommand{\calL}{\mathcal{L}}

\newcommand{\calT}{\mathcal{T}}

\DeclareMathOperator*{\argmax}{argmax}

\theoremstyle{plain}

\theoremstyle{definition}

\algdef{SE}[DOWHILE]{Do}{doWhile}{\algorithmicdo}[1]{\algorithmicwhile\ #1}%

\newcommand{\squishlist}{
 \begin{list}{$\bullet$}
  { \setlength{\itemsep}{0pt}
     \setlength{\parsep}{3pt}
     \setlength{\topsep}{3pt}
     \setlength{\partopsep}{0pt}
     \setlength{\leftmargin}{1.5em}
     \setlength{\labelwidth}{1em}
     \setlength{\labelsep}{0.5em} } }
     
\newcommand{\squishend}{
  \end{list}  }

\newcommand{\name}{\texttt{BOSON}$^{-1}$\xspace}

\def\BibTeX{{\rm B\kern-.05em{\sc i\kern-.025em b}\kern-.08em
    T\kern-.1667em\lower.7ex\hbox{E}\kern-.125emX}}

\begin{document}

\title{BOSON$^{-1}$: Understanding and Enabling Physically-Ro\underline{b}ust Ph\underline{o}tonic \underline{Inv}erse Design with Adaptive Variation-Aware \underline{S}ubspace \underline{O}ptimization 
}

\author
{
Pingchuan Ma$^1$,
Zhengqi Gao$^2$,
Amir Begovic$^3$,
Meng Zhang$^3$,
Haoyu Yang$^4$,
Haoxing Ren$^4$,\\
Rena Huang$^3$,
Duane S. Boning$^2$,
Jiaqi Gu$^1$
\\
$^1$Arizona State University, $^2$MIT, $^3$RPI, $^4$Nvidia\\
\small\textit{\{pingchua, jiaqigu\}@asu.edu}
}

\maketitle
\begin{abstract}
\label{abstract}
Nanophotonic device design aims to optimize photonic structures to meet specific requirements across various applications. 
Inverse design has unlocked non-intuitive, high-dimensional design spaces, enabling the discovery of compact, high-performance device topologies beyond traditional heuristic or analytic methods. 
The adjoint method, which calculates analytical gradients for all design variables using just two electromagnetic simulations, enables efficient navigation of this complex space.
However, many inverse-designed structures, while numerically plausible, are difficult to fabricate and highly sensitive to physical variations, limiting their practical use. 
The discrete material distributions with numerous local-optimal structures also pose significant optimization challenges, often causing gradient-based methods to converge on suboptimal designs.
In this work, we formulate inverse design as a fabrication-restricted, discrete, probabilistic optimization problem and introduce \name, an end-to-end, adaptive, variation-aware subspace optimization framework to address the challenges of manufacturability, robustness, and optimizability.
We explicitly consider the fabrication process and differentiably optimize the design in the fabricable subspace.
To overcome optimization difficulty, we propose dense target-enhanced gradient flows to mitigate misleading local optima and introduce a conditional subspace optimization strategy to create high-dimensional tunnels to escape local optima.
Furthermore, we significantly reduce the prohibitive runtime associated with optimizing across exponential variation samples through an adaptive sampling-based robust optimization method, ensuring both efficiency and variation robustness.
On three representative photonic device benchmarks, our proposed inverse design methodology \name delivers fabricable structures and achieves the best convergence and performance under realistic variations, outperforming prior arts with 74.3\% post-fabrication performance.
We open-source our codes at \href{https://github.com/ScopeX-ASU/BOSON}{link}.
\end{abstract}

\section{Introduction}
Integrated photonics has shown a wide range of applications in computing, communication, and sensing. 
Currently, many photonic devices are manually architected by tuning a few design parameters via inefficient trial and error, which relies heavily on expert knowledge and time-consuming simulations.
In contrast, inverse design requires minimal physical prior knowledge and opens up non-intuitive, high-dimensional design spaces, making it possible to discover highly efficient and compact device designs~\cite{SIM_ACSP2018_Hughes, minkov2020inverse}. 
The adjoint method-based inverse design~\cite{SIM_ACSP2018_Hughes} is particularly powerful for its ability to compute analytical gradients of an objective with respect to high-dimensional design variables using only two simulations. 

While the adjoint inverse design can produce numerically plausible designs, a significant gap exists between pre-fab and post-fab performance.
\begin{figure}
    \centering
    \includegraphics[width=0.75\columnwidth]{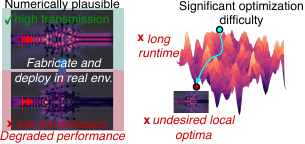}
    \vspace{-5pt}
    \caption{Inverse design often yields non-fabricable devices.
    Optimization difficulty leads to suboptimal designs.}
    \label{fig:motivation}
    \vspace{-5pt}
\end{figure}

As illustrated in Fig.~\ref{fig:motivation}, the inverse-optimized design exhibits high performance, but the tiny structures within the design pattern are \textbf{non-manufacturable}, leading to severe post-fabrication performance degradation. 
Additionally, \textbf{fabrication variations} during lithography and etching, along with \textbf{operational variations} such as temperature drift, introduce robustness concerns that further compromise the performance of inverse-designed devices.
Previous approaches have attempted to address manufacturability issues by controlling the minimum feature size (MFS) during optimization through heuristic methods, such as blurring or adding curvature penalties, to eliminate non-fabricable structures and mitigate post-fabrication performance drops~\cite{DID_ACSP2020_Chen, DID_NP2023_Mao, DID_ACSP_Schubert, b-spline, sigmud_density, wang2019robust, spatial_varying, zhou2024inverse, gershnabel2022reparameterization, hammond2022high}.
To enhance fabrication robustness, prior work models variations as uniform erosion and dilation of the device geometry and simultaneously optimizes objectives under different variation corner cases. 
However, this oversimplified method only marginally improves robustness as it fails to capture actual variations accurately.
When considering multiple variation effects, exhaustive Monte Carlo sampling of all corner cases induces \textbf{exponential simulation cost}.
Beyond fabricability, solving this \textbf{discrete high-dimensional stochastic optimization problem is particularly challenging}.
As the adjoint optimizer is only driven by a single objective sparsely defined in the device output port, it leads to a \textbf{poor objective landscape} shown in Fig.~\ref{fig:motivation}, making the optimization highly sensitive to initialization and prone to getting trapped in unreasonable suboptimal solutions.

To address the photonic device inverse design challenge, which requires both fabricability and robustness, we formulate this task as a fabrication-restricted, robust stochastic optimization problem. 
We propose a novel inverse design framework, \name, which enables effective optimization directly within the fabricable subspace with full variation awareness, ensuring efficient inverse design toward robust device structures. 
Our main contributions are as follows:
\squishlist
    {\item We provide a comprehensive analysis of the fabricability and optimization challenges in inverse design and introduce \name, an adaptive, variation-aware inverse design framework for physically robust photonic devices.
    }
    {\item \textbf{Fabrication-Aware Subspace Optimization}: \name integrates differentiable fabrication modeling into adjoint optimization to ensure devices in the fabricable subspace.
    }
    {\item \textbf{Loss Landscape Reshaping}: \name largely reduces the specious local optima by introducing auxiliary dense objectives, enhancing gradient flow and improving optimization.
    }
    {\item \textbf{Local Optima Escaping}: We introduce a light-concentrated initialization method with conditional subspace relaxation, facilitating escaping local optima toward better solutions.
    }
    {\item \textbf{Linear-Cost Variation-aware Optimization}: We propose an adaptive variation optimization method based on novel axial corner sampling and worse-case optimization, reducing simulation costs from exponential to linear.
    }
    {\item On three photonic device benchmarks, our \name achieves 74.3\% performance enhancement on average compared to previous art, enabling variation-robust photonic inverse design with high efficiency.}
\squishend

\section{Preliminary}
\label{sec:Background}
We will briefly illustrate the basic concepts in fabrication and variations and review previous work focusing on improving fabricability and robustness.
\subsection{Fabrication and Operation Variations}
\label{sec:fabVarIntro}
The binary photonic device pattern from inverse design often deviates from the real device due to the limited resolution in fabrication and the non-ideality in actual operation, e.g., temperature drift. 
We decoupled this complication process into the following three steps as shown in Fig.~\ref{fig:fabConstrainedVarAware}.

\noindent\textbf{Lithography}.~
During photolithography, light shines through the mask to transfer the design pattern onto the wafer, as shown in step 1 of Fig.~\ref{fig:fabConstrainedVarAware}. However, due to diffraction, the light intensity distribution on the wafer differs from the intended design, often leading to the loss of small features. Moreover, lithography in practice induces variations in focus, defocus, and exposure dose, causing further structure discrepancy from the desired one. 
In this work, we utilize a differentiable lithography model\cite{DID_ICML2024_Yang} based on the Hopkins diffraction equation to predict the three corner cases of the post-lithography pattern using the inverse-designed mask.

\noindent\textbf{Etching}.~Following lithography, the etching process is applied to the exposed wafer, removing material from the exposed regions to form the device structures.
We model this process as a binarization projection applied to the continuous post-lithography pattern using a specific threshold, $\eta$. 
In practice, etching introduces spatial variations across the wafer. 
In this work, we model these spatially varying etching effects by characterizing $\eta$ as a random field, utilizing the expansion optimal linear estimation (EOLE) method~\cite{DID_CMAME2011_Sche}.

\noindent\textbf{Operation Variations}.~
The last step in Fig.~\ref{fig:fabConstrainedVarAware} refers to the variation during practical operation that will cause the permittivity to drift from its nominal value. 
In this work, we consider the temperature-caused permittivity drift. Since we use air as the void (cladding) and silicon as the solid to build the photonic devices, we model the temperature-dependent Si permittivity by $\epsilon_{Si}(t) = (3.48 + 1.8\times10^{-4}\cdot(t - 300))^2$~\cite{DV_APL2012_Komma}

\subsection{Prior Fabrication-Robust Photonic Inverse Design Methods}
Minimum feature size (MFS) control is a widely used technique to enhance fabricability. 
Designs can be intentionally parametrized to form smooth gating structures~\cite{DID_ACSP2020_Chen} that meet the MFS requirements of the foundry.
However, this method is not applicable to complex 2-D patterns.
Low-pass filters or blurring have been applied to remove high-frequency features from pre-fabrication patterns~\cite{DID_NP2023_Mao}.
This method is only an approximation of the lithography process and cannot fully guarantee performance after actual lithography and etching.
To account for etching variations, previous work has optimized both the nominal design and the dilated/eroded versions to achieve robust gating structures against over-etching and under-etching~\cite{DID_ACSP2020_Chen, hammond2021photonic, wang2011robust}. 
However, the uniform etching threshold used in this method oversimplifies the actual etching effects, which limits their robustness in real-world fabrication scenarios.
Moreover, the exhaustive corner sweeping-based method~\cite{DID_ACSP2020_Chen} results in exponential simulation costs, making it unscalable for handling more complex variation models.

\section{Robust Photonic Inverse Design with Adaptive Variation-Aware Subspace Optimization}
We first give a formulation of robust inverse design problem and analyze its fabrication constraints and robust optimization challenges. 
To solve those challenges, \name proposes to directly optimize the design pattern in \textbf{a low-dimensional manufacturable subspace} and enable efficient and robust optimization via gradient-enhanced landscape reshaping, conditional subspace optimization, and adaptive variation sampling. 

\subsection{Problem Formulation}
\label{subsec:formulation}
Our \textbf{goal} is to optimize the design variables $\theta \in \mathbb{R}^{N}$ so that the corresponding binary design pattern $\epsilon$ can obtain the maximum expected post-fabrication figure-of-merits (FoM) under variations. 
We formulate this task as a \textbf{constrained, discrete, stochastic} optimization problem as follows,
\begin{equation}
\small
\label{eq:Formulation}
\begin{gathered}
\theta^* = \argmax_{\theta \in \Theta}~~\mathbb{E} _{T, \eta, L}F(\epsilon(\theta)|\lambda_c), \\
\text{s.t.}~~
\Tilde{\Bar{\rho^{\prime}}} = (\mathcal{T}_t\circ \mathcal{E}_{\eta}\circ \mathcal{L}_l\circ \mathcal{P})(\theta)\\ 
\epsilon = \epsilon_v + (\epsilon_s-\epsilon_v)\cdot \Tilde{\Bar{\rho^{\prime}}}\\
\mathcal{P}: \theta \in \mathbb{R}^N \to \rho \in \{0, 1\}^{N^x \times N^y}\\
\mathcal{L}_l: \rho \in \{0, 1\}^{N^x \times N^y} \to \Bar{\rho} \in [0, 1]^{N^x \times N^y} \\
\mathcal{E}_{\eta}: \Bar{\rho} \in [0, 1]^{N^x \times N^y} \to \Tilde{\Bar{\rho}} \in \{0, 1\}^{N^x \times N^y} \\
\mathcal{T}_t: \Tilde{\Bar{\rho}} \in \{0, 1\}^{N^x \times N^y} \to \Tilde{\Bar{\rho^{\prime}}} \in \{0, \alpha_t\}^{N^x \times N^y}\\
\end{gathered}
\end{equation}
where $\theta$ are latent design variables that encode the topology of the device pattern $\epsilon$.
The compound mapping function from $\theta$ to $\epsilon$ includes the following transformations.
(1) $\mathcal{P}$: it maps $\theta$ to a Boolean pattern $\rho$ that indicates the probability of solid material $\epsilon_s$ and void space $\epsilon_v$. In this work, we adopt a popular level set method~\cite{DID_CMAME2003_Wang} to parameterize the topology.
(2) $\mathcal{L}_l$: the lithography model maps $\rho$ to a non-binary post-litho pattern $\Bar{\rho}$;
(3) $\mathcal{E}_{\eta}$: the etching model maps the post-litho pattern to a binarized post-etching pattern $\Tilde{\Bar{\rho}}$;
and (4) $\calT_t$: it represents the actual operational condition that can potentially change the device permittivity.

The target is not to maximize the objective $F(\cdot)$ for central wavelength $\lambda_c$ at a single nominal device design point, but the \textbf{expected performance over random variations during fabrication and operation} $T$, $\eta$, and $L$.
$L \in \{l_{min}, l_{norm}, l_{max}\}$ refers to different lithography corners due to defocusing.
$\eta$ is the random spatially varying etching threshold field that leads to pattern distortion.
$T$ represents the undesired permittivity drift during actual device operation, e.g., temperature drift.

\subsection{Understanding the Fabricability Constraint and Optimization Difficulty of Photonic Inverse Design}
Based on the above complex formulation, we thoroughly analyze its \emph{subspace optimization property and pinpoint sources of optimization difficulty}, which motivates our proposed \name methodology.

\begin{figure*}
    \centering
    \subfloat[]{\includegraphics[width=0.45\textwidth]{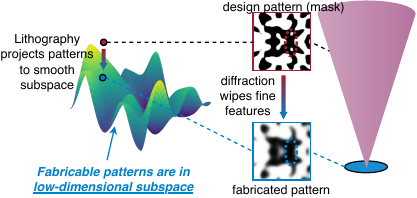}
    \label{fig:FabIllustration}
    }
    \hspace{10pt}
    \subfloat[]{\includegraphics[width=0.4\textwidth]{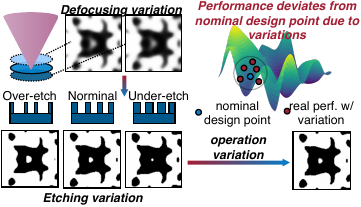}
    \label{fig:VariationIllustration}
    }
    \caption{(a) Lithography and etching during fabrication restricts manufacturable patterns in the subspace.
    (b) Fabrication variations (defocusing and etching) and operation variations cause performance deviation.}
    \label{fig:fabConstrainedVarAware}
    \vspace{-10pt}
\end{figure*}

\subsubsection{Fabrication Constraints}
We decouple the fabrication into two cascaded processes: lithography $\mathcal{L}_l$ and etching $\mathcal{E}_{\eta}$.

\noindent\textbf{Lithography}.~As shown in Fig.~\ref{fig:FabIllustration}, fine-grained voids or solids are blurred and removed, and edges with shape curvature are smoothed since the feature sizes are smaller than the diffraction limit.
This is the main mechanism that restricts the device patterns to the low-dimensional manufacturable subspace.

\noindent\textbf{Etching}.~As shown in Fig.~\ref{fig:VariationIllustration}, etching is modeled as a binarization projection of the post-litho patterns. 
Small features can hardly survive under the non-ideal etching threshold, e.g., under-etching removes small holes in Fig.~\ref{fig:VariationIllustration}.

Both processes have non-ideal variations in practice, which complicates the inverse design problem as a stochastic optimization over different fabrication conditions.

\subsubsection{Optimization Difficulty}
\label{sec:optDiff}
\begin{figure*}
    \centering
    \includegraphics[width=1\textwidth]{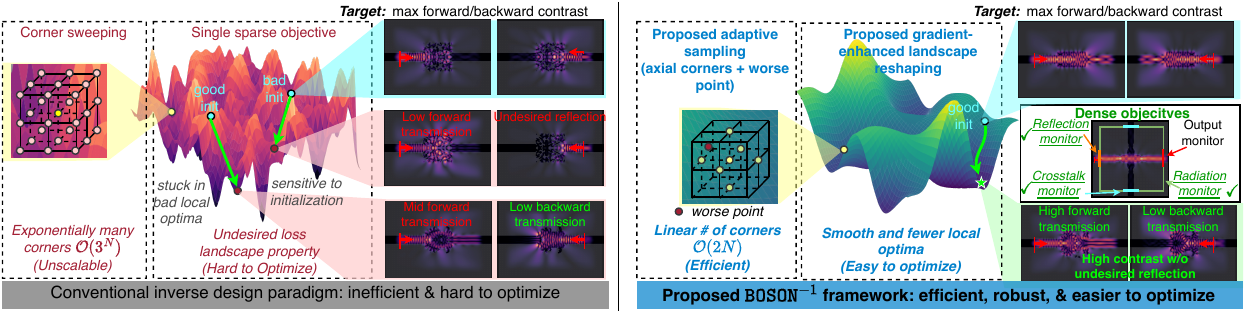}
    \caption{Our proposed \name framework enables efficient, robust optimization with better convergence and optimality.}
    \label{fig:optDiff}
    \vspace{-10pt}
\end{figure*}

A common observation of photonic inverse design is that the adjoint optimization often converges to \textbf{sub-optimal or even unreasonable design patterns}.
We attribute this phenomenon to the following reasons.
\ding{202}~The traditional single objective, e.g., transmission, leads to poor properties of the loss landscape, where \textbf{numerous undesired yet sharp local optima/saddle points} hinder effective exploration.
\ding{203}~The traditional \textbf{design objective is overly sparse}, i.e., transmission through a small monitor, which causes optimization instability due to vanishing gradients when no lights pass through the monitor. 
\ding{204} Due to the binary nature of the optimization, the \textbf{highly discrete} non-convex problem leads to high \textbf{sensitivity to initialization}.
\ding{205}~Stochastic optimization requires optimizing over the entire variation distribution.
Naive Monte Carlo sampling or corner sweeping method has \textbf{a prohibitive cost and attention distraction} and is thus unstable, inefficient, and leads to sub-optimal robust design.

To tackle these difficulties, several questions have to be answered first:

\noindent\ding{202}~\textit{What defines a good loss landscape for adjoint gradient-based photonic inverse design?} Finding a high-quality design becomes exceedingly challenging with a poorly structured loss landscape. 
In Fig.~\ref{fig:optDiff}, we illustrate this using an optical isolator as an example, where the objective is to minimize transmission contrast ratio $\frac{E_{bwd}}{E_{fwd}}$ by maximizing forward transmission and minimizing backward transmission. 
This objective leads to a poor loss landscape where the optimizers are easily stuck at specious designs.
For example, from a randomly initialized pattern, it converges to a low contrast result by reducing the backward transmission, even when the forward transmission is unreasonably low. 
Even with a good heuristic initialization, it converges to a sub-optimal point with unsatisfying $E_{fwd}$.
The objective fails to guide the optimization to a desired space.
Hence, it is essential to \textbf{reshape the loss landscape with fewer sharp yet undesired local optima}.

\noindent\ding{203}~\textit{How can we ensure a richer gradient flow?} As shown in Fig.~\ref{fig:optDiff}, the gradient flow is weak when the optimization objective is based solely on the power collected at the output port, widely adopted in photonic inverse design. 
Drawing an analogy to the gradient vanishing problem in neural networks, the optimization stagnates when all neurons are dead, e.g., due to ReLU activations. 
In the case of photonic inverse design, without careful initialization, strong local reflection and radiation occur easily in early optimization stages, leading to no light passing through the output monitor and thus vanishing gradients. 
This, in turn, traps the optimization in sub-optimal or unreasonable design points. 
We conclude that inverse design based on sparse objective is \textbf{inherently unstable from the gradient's perspective}. 
To mitigate this, inspired by dense supervision in neural network knowledge distillation\cite{NN_ICLR2015_Romero}, we are compelled to \textbf{provide \emph{dense supervision} with auxiliary objectives to ensure a richer gradient flow, equivalently smoothing out the loss landscape}, preventing the optimization from stagnating in unsuitable patterns.

\noindent\ding{204}~\textit{How can we efficiently sample the variation distribution for robust optimization?} 
Monte Carlo sampling is the common method to estimate the behavior of a distribution.
As a heuristic, hardware designers tend to use min-max sampling to sweep all variation corners.
However, as shown in Fig.~\ref{fig:optDiff}, the \textbf{exponentially many variation corner cases make it unscalable} for this simulation-in-the-loop inverse design. 
To address this, it is crucial to draw samples adaptively, ideally with linear cost, to \textbf{balance optimization efficiency and variation robustness}.

\noindent\ding{205}~\emph{How to initialize the design?} The discrete nature of the binary pattern design problem makes it sensitive to initialization.
A randomly initialized pattern tends to scatter lights to different paths, which biases the converged pattern to an undesired design with many fabrication-unfriendly small structures.
Besides, this can lead to a physically unstable local resonance state where the optimizer can hardly escape.
To solve this problem, it is most effective if a \textbf{good heuristic initialization} can be found.
Furthermore, it is preferred to reduce the initialization sensitivity by having a smooth loss landscape with effective optimization strategies to \textbf{escape from local optima}.

\subsection{Fabrication-Restricted Subspace Optimization Toward Guaranteed Manufacturability}
To minimize the post-fab performance gap in conventional methods using heuristic MFS control and a separate inverse lithography mask correction step\cite{prefab, gao2014mosaic}, \name enables \emph{end-to-end inverse design with explicit fabrication and variation modeling} in the optimization loop.

We adopt a Hopkins diffraction-based lithography modeling\cite{DID_ICML2024_Yang} $\calL(\cdot)$ and a gradient-estimated etching modeling during device construction before evaluating the adjoint gradient to enable differentiable fab-aware topology optimization.
This \emph{re-parametrization} trick enforces the generated pattern to be aware of realistic smoothing and distortion from the actual fabrication modeling, such that the optimization trajectory is restricted to the manufacturable subspace.
Compared to mask correction, which tries to match the freely optimized pattern at the post-processing stage, our \textbf{subspace optimization method guarantees fabricability and eliminates the performance gap due to topology discrepancy} after correction.

\subsection{Gradient-Enhanced Subspace Adjoint Optimization Toward Better Optimality}
We solve the optimization difficulty of photonic inverse design from three aspects: (1) improve the loss landscape property, 
(2) better method to escape local optima, and 
(3) better initialization.
\subsubsection{Gradient-Enhanced Objective Landscape Reshaping via Auxiliary Constraints}
As discussed in Section~\ref{sec:optDiff}, a sparse objective leads to numerous local optima that hinder optimization.
Hence, we guide the optimization with auxiliary constraints $F_i(\epsilon|\lambda_c) \leq C_i$ and relax the inequality constraints as additional penalty terms in the objective as follows,
\begin{equation}
    \small
    \label{eq:Penalty}
    obj = F(\epsilon| \lambda_c) + \sum _iw_i\cdot [F_i(\epsilon| \lambda_c) - C_i]_+,
\end{equation}
where $(\cdot)_+$ adds penalty only when constraints are violated.
For the isolator design as an example, we encourage forward transmission higher than 80\%, reflection less than 10\%, backward radiation higher than 90\%, etc.
At the early stage, the penalty terms calculated on extra power monitors, as shown in Fig.~\ref{fig:optDiff}, provide dense supervision and enhanced gradient flow that push the design towards a desired region.
Bad and sharp local optima in the original loss landscape disappear in the reshaped design space.
Moreover, gradient vanishing problems when using a small output port monitor are resolved as the added auxiliary monitors provide rich gradients for optimization.
As the penalty terms disappear once constraints are satisfied, the optimization is driven by the main objective, eventually converging to a high-FoM and feasible design.

\subsubsection{Subspace Local Optima Escaping via Relaxed High-Dimensional Tunneling}
The main reason that the variation-aware optimization is easier to get stuck in the local optima is that the optimization is restricted in a low-dimensional fabricable subspace which is harder for the optimizer to escape sharp local optima. 
Furthermore, the gradients backpropagated through the lithography model are hurdled as the model filters out small features, which cause gradient vanishing on small structures. \begin{figure}
    \centering
    \vspace{-5pt}
    \includegraphics[width=0.95\columnwidth]{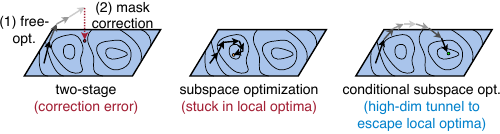}
    \vspace{-5pt}
    \caption{High-dimensional tunnel for local minima escaping.}
    \label{fig:subspaceRexIllustration}
    \vspace{-10pt}
\end{figure}

So instead of solely relying on the gradient from projected design pattern $\Tilde{\Bar{\rho^{\prime}}} = (\mathcal{T}_t\circ \mathcal{E}_{\eta}\circ\mathcal{L}_l\circ\mathcal{P})(\theta)$, we add a high-dimensional tunnel outside the fabricable subspace through pattern $\rho$ as shown in Fig.~\ref{fig:subspaceRexIllustration}, making the optimization objective a weighted sum of fabrication-aware and ideal term,  
\begin{equation}
\small
    obj = p \cdot \mathbb{E}_{T, \eta, L} \left[ \sum_i w_i F_i\left( \widetilde{\overline{\epsilon'}}(\theta)| \lambda_c \right) \right] \\
          \!+ (1 - p) \cdot \sum_i w_i F_i\left( \epsilon(\theta)| \lambda_c \right).
\end{equation}
$p$ will gradually increase to 1 to ensure variation awareness and guarantee fabricability. 
With this technique, the subspace optimization can easily escape from the local optima.
\subsubsection{Optical Path Concentrated Initialization}
A well-chosen initialization helps prevent the optimization process from getting trapped in suboptimal points. In \name, rather than randomly initializing the design parameters $\theta$, which can lead to undesired initial optical behavior or even significant reflection that hinders optimization as discussed in Section~\ref{sec:optDiff}, we initialize the design region to simple yet effective geometry with concentrated optical paths to ensure enough light passing through monitors with strong gradients.

\subsection{Adaptive Sampling Strategy for Efficiency and Robustness}
The number of corners increases exponentially with the number of parameters characterizing the fabrication error, which makes the prior exhaustive sampling method unscalable when considering multiple error sources. 
For example, in our error modeling, there are 3 lithography corners, 3 temperature corners, and 3 global etching thresholds if we consider a simpler etching model. 
In total, there are $3^3 = 27$ corners, requiring$54$ simulations in one iteration. 
This motivates us to adopt an adaptive sampling strategy. 
Besides the nominal corner $C_n$, we sample 6 more corners, namely, $C_L$, $C_{T}$ and $C_{\eta}$ in which $C_L$ represents the minimum and maximum corner of the lithography model and use the nominal value for the temperature and global etching threshold $\eta$ and so does the rest two as shown in Fig.~\ref{fig:optDiff}. 
Based on this \textbf{axial sampling strategy}, we \textbf{reduce the complexity from exponential to linear}. 
However, these sampled corners still cannot model the spatially varying etching field. 
As such, inspired by the robust optimization on neural network\cite{foret2021sharpnessaware, pmlr-v139-kwon21b}, we sample an additional worst corner $C_{worst}$ by one-step gradient ascend on the temperature and the weight of the etching field basis. 
Combining axial corners and worst corners, we can ensure both efficiency and robustness.

\section{Evaluation Results}
\label{sec:Experiment}
\subsection{Evaluation Setup}
\paragraph{Benchmarks}
We adopt three representative photonic devices to demonstrate the effectiveness and efficiency of \name: (1) a waveguide bending, where light is steered by $90^\circ$; (2) a waveguide crossing, where light propagates through intersecting waveguides without crosstalk; and (3) an optical isolator, where light is converted from $TM_1$ mode to $TM_3$ mode in forward propagation with high transmission efficiency while the backward light is isolated via radiation.

\begin{figure*}
    \centering
    \subfloat[]{\includegraphics[width=0.32\textwidth]{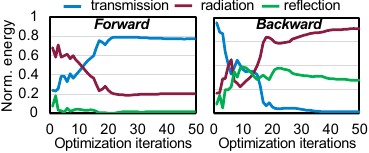}
    \label{fig:trajMultiFoM}
    }
    \hspace{3pt}
    \subfloat[]{\includegraphics[width=0.32\textwidth]{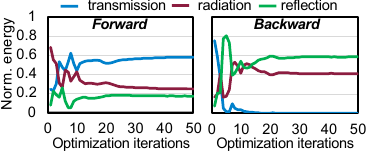}
    \label{fig:trajSingleFoMGoodInit}
    }
    \hspace{3pt}
    \subfloat[]{\includegraphics[width=0.31\textwidth]{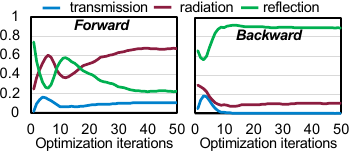}
    \label{fig:trajSingleFoMBadInit}
    }
    \vspace{-5pt}
    \caption{Fabrication-aware optimization trajectories of optical isolator with forward/backward transmission, radiation, and reflection.
    No variation is added.
    (a) Proposed method (light-concentrated initialization, dense objectives).
    (b) Light-concentrated initialization with contrast as the single sparse objective.
    (c) Random initialization with contrast as the single sparse objective.
    }
    \label{fig:optTrajectory}
    \vspace{-10pt}
\end{figure*}

\paragraph{Objectives} The optimization objectives are transmission efficiency for the bending and the crossing $\frac{E_{out}}{E_{in}}$. 
For the optical isolator, the objective is the isolation contrast $\frac{E_{bwd}}{E_{fwd}}$ where $E_{bwd}$ and $E_{fwd}$ refer to the transmission efficiency in backward and forward propagation.

\subsection{Main Results}

\begin{table}[]
\centering
\caption{Main result, the FoM for bending and crossing is the transmission efficiency, the higher the better; the FoM for isolator is the isolation contrast, the lower the better}
\resizebox{1\columnwidth}{!}{%
\begin{tabular}{|cccc|}
\hline
\multicolumn{1}{|c|}{Benchmark}                            & \multicolumn{1}{c|}{Model}                                 & \multicolumn{1}{c|}{Fwd \& bwd transmission}                               & Avg FoM                                  \\ \hline
\multicolumn{1}{|c|}{}                                     & \multicolumn{1}{c|}{Density}                               & \multicolumn{1}{c|}{N/A}                                                   & 0.916 → 0.0487                           \\ \cline{2-4} 
\multicolumn{1}{|c|}{}                                     & \multicolumn{1}{c|}{InvFabCor-M-3}                         & \multicolumn{1}{c|}{N/A}                                                   & 0.953 → 0.7                              \\ \cline{2-4} 
\multicolumn{1}{|c|}{}                                     & \multicolumn{1}{c|}{\cellcolor[HTML]{EFEFEF}\textbf{\name}} & \multicolumn{1}{c|}{\cellcolor[HTML]{EFEFEF}\textbf{N/A}}                  & \cellcolor[HTML]{EFEFEF}\textbf{0.967}   \\ \cline{2-4} 
\multicolumn{1}{|c|}{\multirow{-4}{*}{Crossing}}           & \multicolumn{3}{c|}{avg   improvement: 60\%}                                                                                                                                       \\ \hline
\multicolumn{1}{|c|}{}                                     & \multicolumn{1}{c|}{Density}                               & \multicolumn{1}{c|}{N/A}                                                   & 0.996 → 0.0141                           \\ \cline{2-4} 
\multicolumn{1}{|c|}{}                                     & \multicolumn{1}{c|}{InvFabCor-M-3}                         & \multicolumn{1}{c|}{N/A}                                                   & 0.935 → 0.691                            \\ \cline{2-4} 
\multicolumn{1}{|c|}{}                                     & \multicolumn{1}{c|}{\cellcolor[HTML]{EFEFEF}\textbf{\name}} & \multicolumn{1}{c|}{\cellcolor[HTML]{EFEFEF}\textbf{N/A}}                  & \cellcolor[HTML]{EFEFEF}\textbf{0.982}   \\ \cline{2-4} 
\multicolumn{1}{|c|}{\multirow{-4}{*}{Bending}}            & \multicolumn{3}{c|}{avg   improvement: 63\%}                                                                                                                                       \\ \hline
\multicolumn{1}{|c|}{}                                     & \multicolumn{1}{c|}{Density}                               & \multicolumn{1}{c|}{{[}0.676, 3.53e-06{]} →   {[}0.0204, 0.0757{]}}        & 4.89E-06 → 3.71                          \\ \cline{2-4} 
\multicolumn{1}{|c|}{}                                     & \multicolumn{1}{c|}{InvFabCor-M-3}                         & \multicolumn{1}{c|}{{[}0.152, 2.361e-4{]} → {[}0.0228, 0.0275{]}}          & 0.00156 → 0.528                          \\ \cline{2-4} 
\multicolumn{1}{|c|}{}                                     & \multicolumn{1}{c|}{\cellcolor[HTML]{EFEFEF}\textbf{\name}} & \multicolumn{1}{c|}{\cellcolor[HTML]{EFEFEF}\textbf{{[}0.8275, 0.0022{]}}} & \cellcolor[HTML]{EFEFEF}\textbf{0.00262} \\ \cline{2-4} 
\multicolumn{1}{|c|}{\multirow{-4}{*}{Optical   isolator}} & \multicolumn{3}{c|}{avg   improvement: 100\%}                                                                                                                                      \\ \hline
\multicolumn{4}{|c|}{total avg   improvement: 74.3\%}                                                                                                                                                                                           \\ \hline
\end{tabular}
}
\label{tab:mainResult}
\end{table}

\noindent\textbf{Benchmark Notations}.~Before presenting the results, we define the baseline notations. 'LS' and 'Density' denote level set or density-based methods for parameterizing the design, with \name using level set by default. 'InvFabCor' refers to a two-stage process: first, levelset is used to achieve a high-performance design $\rho^*$, followed by mask optimization to match the post-fabrication pattern. '-M' indicates MFS control is included, while '-\#' specifies the number of lithography corner to match. '-eff' signifies that the objective is transmission efficiency, not isolation contrast used in optical isolator. The arrow ($\rightarrow$) in the tables shows performance transitioning from the optimized result to the evaluated real-world performance.

To evaluate the performance of different design patterns produced by various optimization methods, we use Monte Carlo sampling, where lithography corners, random $\eta$ fields, and temperature are treated as random variables. We measured the average of 20 samples for each case under uniform distribution, and the result is shown in Table~\ref{tab:mainResult}.
In all three benchmarks, traditional density-based inverse design produces numerically plausible results. However, without MFS control, these designs contain too many fine features that are non-fabricable, resulting in poor post-fabrication performance. While inverse fabrication correction finds high-performance designs near the fabricable space, performance degrades sharply after projection into the fabricable subspace, highlighting the need for direct optimization within this subspace. In the optical isolator benchmark, inverse fabrication correction failed to produce a viable device due to poor initialization. In contrast, \name achieves the best performance across all benchmarks by directly optimizing the fabricable subspace, using well-chosen initialization, and applying dense supervision.
\subsection{Ablation Study and Discussion}
In the ablations study and discussion section, all the experiment results are obtained on optimizing the most challenging benchmark, the optical isolator, where various effects, such as reflection, resonance, and radiation, are involved.
\paragraph{Ablation Study}\begin{table}[]
\centering
\caption{Ablation study of \name}
\resizebox{1\columnwidth}{!}{%
\begin{tabular}{|c|c|c|c|}
\hline
model                      & {[}fwd, bwd{]}                & contrast ↓        & degradation \\ \hline
\rowcolor[HTML]{EFEFEF} 
\textbf{\name}              & \textbf{{[}0.8275, 0.0022{]}} & \textbf{2.62E-03} & \textbf{N/A}         \\ \hline
- loss landscape reshaping & {[}0.4382, 0.0023{]}          & 5.41E-03          & 52\%                 \\ \hline
- subspace relax           & {[}0.8066, 0.0025{]}          & 3.25E-03          & 19\%                 \\ \hline
exhaustive   sample        & {[}0.8395, 0.0113{]}          & 1.36E-02          & 81\%                 \\ \hline
random init                & {[}0.0610, 0.0356{]}          & 7.04E-01          & 100\%                \\ \hline
\end{tabular}
}
\label{tab:AblationStudy}
\end{table}
 
Table~\ref{tab:AblationStudy} compares our method with its variants, each lacking a key technique. Without loss landscape reshaping and dense supervision, contrast performance decreases by 52\%, and more critically, the forward efficiency is severely compromised. Omitting subspace relaxation results in a 19\% reduction in contrast, replacing adaptive sampling with exhaustive sampling leads to a dramatic 81\% decrease in contrast, and most critically if a random initialization is used, the device becomes invalid.

\paragraph{Loss Landscape Reshaping and Dense Objectives} 
As shown in Fig.~\ref{fig:optTrajectory}, without loss landscape reshaping, random initialization causes forward transmission efficiency to stagnate due to vanishing gradients. 
While the backward efficiency is low, giving a seemingly good contrast ratio, this is mainly due to unwanted reflection. 
Even with better initialization, the poor loss landscape results in low backward efficiency and suboptimal forward efficiency. 
In contrast, reshaping the loss landscape and using dense objectives with multiple penalties guide the design to a high-performance solution with strong isolation contrast and good functionality.

\paragraph{Subspace Relaxation}

\begin{figure}
    \centering
    \vspace{-5pt}
    \subfloat[]{\includegraphics[width=0.54\columnwidth]{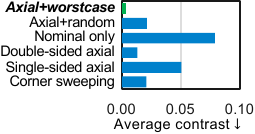}
    \label{fig:sampleMethod}
    }
    \hspace{0pt}
    \subfloat[]{\includegraphics[width=0.41\columnwidth]{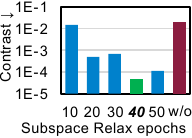}
    \label{fig:subspaceRelax}
    }
    \caption{
    (a) Comparison between different sampling strategies, Axial + worst case shows best average contrast;
    (b) Comparison between different relaxation epochs and no relaxation.
    }
\label{fig:sampleAndSubspaceRex}
\end{figure}

\begin{table}[]
\centering
\caption{Compare different methods for optical isolator}
\resizebox{1\columnwidth}{!}{%
\begin{tabular}{|c|c|c|}
\hline
model             & Fwd \& bwd   transmission                  & Avg FoM          \\ \hline
Density           & {[}0.625, 4.14e-4{]}→{[}0.419, 0.298{]}    & 3.7e-4→0.763     \\ \hline
Density-M         & {[}0.405, 0.275{]}→{[}0.0883, 0.737{]}     & 0.674→9.83       \\ \hline
LS                & {[}0.0547, 0.00137{]}→{[}0.0126, 0.0339{]} & 0.0252→2.91      \\ \hline
LS-M              & {[}0.722, 0.00140{]}→{[}0.604, 0.400{]}    & 0.00208→0.677    \\ \hline
InvFabCor-1       & {[}0.0547, 0.00138{]}→{[}0.0398, 0.0163{]} & 0.0252→0.563     \\ \hline
InvFabCor-3       & {[}0.0547, 0.00138{]}→{[}0.0335, 0.0162{]} & 0.0252→0.627     \\ \hline
InvFabCor-M-1     & {[}0.722, 0.00150{]}→{[}0.684, 0.0201{]}   & 0.00208→0.0306   \\ \hline
InvFabCor-M-3     & {[}0.722, 0.00150{]}→{[}0.687, 0.0179{]}   & 0.00208→0.0271   \\ \hline
InvFabCor-M-3-eff & {[}0.990,   0.669{]}→{[}0.891, 0.674{]}    & 0.676→0.756      \\ \hline
\rowcolor[HTML]{EFEFEF} 
\textbf{\name}     & \textbf{{[}0.828, 0.0022{]}}               & \textbf{0.00262} \\ \hline
\end{tabular}
}
\label{tab:isoCmpr}
\vspace{-10pt}
\end{table}

Figure~\ref{fig:subspaceRelax} shows the optimized contrast under different epochs of relaxation and the one without subspace relaxation, from which it is clear that subspace relaxation can greatly improve the contrast ratio by around 400 times. 
Note that the hyperparameter is searched on the nominal corner without variation.

\paragraph{Adaptive Sampling} 
We evaluate exhaustive sampling ($\mathcal{O}(3^N)$), single-sided axial sampling ($\mathcal{O}(N)$), and axial sampling ($\mathcal{O}(2N)$) in Fig.~\ref{fig:sampleMethod}.
Single-sided axial performs poorly due to its asymmetry, while axial sampling surprisingly outperforms exhaustive sampling by avoiding low-probability corner cases. 
Without accounting for variations (Nominal-only), performance drops significantly. 
Axial sampling with one worst-case corner boosts performance by 6$\times$ compared to two random samples with the same simulation cost.

\paragraph{Compare Different Methods} 
Table~\ref{tab:isoCmpr} compares different methods with good initialization.
Our \name outperforms the strongest baseline (two-stage variation-aware optimization with mask correction) with one order of magnitude better FoM, even with fabrication variations.

\section{Conclusion}
\label{sec:Conclusion}
In this work, we propose \name, a photonic device inverse design framework that addresses the manufacturing constraints by directly optimizing the device patterns in the fabricable subspace.
Various optimization techniques have been integrated to reshape the lass landscape and mitigate the optimization difficulty for more efficient exploration toward variation-robust devices. 
We evaluate \name across three photonic device inverse design tasks, demonstrating an average performance improvement of 74.3\% over prior approaches.


\end{document}